\documentstyle[preprint]{aastex}

\begin{document}

\title{Main-Belt Comet 238P/Read Revisited
        \footnote{
        Some of the data presented herein were obtained at the W. M.
        Keck Observatory, which is operated as a scientific partnership
        among the California Institute of Technology, the University of
        California, and the National Aeronautics and Space Administration.
        The Observatory was made possible by the generous financial support
        of the W. M. Keck Foundation.  Some data presented herein were also
        obtained at ESO facilities at La Silla under program ID 085.C-0363(A).
        }
}

\author{Henry H. Hsieh$^{a,b,c}$, Karen J.\ Meech$^{a}$, \& Jana Pittichov\'a$^{a}$}

\affil{
    $^{a}$Institute for Astronomy, Univ.\ of Hawaii, 2680 Woodlawn Drive, Honolulu, Hawaii 96822, USA\newline
    $^{b}$Astrophysics Research Centre, Queen's University, Belfast, BT7 1NN, United Kingdom\newline
    $^{c}$Hubble Fellow\newline
}
\email{hsieh@ifa.hawaii.edu, meech@ifa.hawaii.edu, jana@ifa.hawaii.edu}

\slugcomment{Submitted, 2011-05-10; Accepted, 2011-05-31}

\begin{abstract}
We present a series of observations of the return of activity in main-belt comet 238P/Read.
Using data obtained in July and August 2010 when 238P appeared to be largely inactive, we find best-fit 
IAU phase function parameters of $H=19.05\pm0.05$~mag, corresponding to a nucleus radius of
$r_n\approx0.4$~km (assuming an albedo of $p_R=0.05$), and $G=-0.03\pm0.05$.
Observations from September 2010 onward show a clear rise in activity, causing both a notable 
change in visible morphology and increasing photometric excesses beyond what would be expected
based on bare nucleus observations.  By the end of the observing period reported on here,
the dust mass in the coma shows indications of reaching a level comparable to that observed in 2005, but further observations
are highly encouraged once 238P again becomes observable from Earth in late 2011 to confirm
whether this level of activity is achieved, or if a notable decrease in activity 
strength compared to 2005 can be detected. 
Comet 238P is now the second main-belt comet (after 133P/Elst-Pizarro) observed to exhibit recurrent activity,
providing strong corroboration for the conclusion that it is a true comet whose
active episodes are driven by sublimation of volatile ice.
\end{abstract}


\newpage

\section{INTRODUCTION}

Main-belt comets (MBCs) are a recently-identified population of solar system objects that
exhibit cometary activity but are dynamically indistinguishable from main-belt
asteroids \citep{hsi06}.  Discovered on UT 2005 October 24 \citep{rea05}, 
comet 238P/Read (hereafter, 238P) was the second main-belt comet to be discovered.
A physical study of the comet shortly following its discovery found that its activity
was consistent with dust emission driven by the sublimation of recently excavated subsurface ice,
and inconsistent with impact-driven dust emission, meaning that 238P is
a true comet \citep{hsi09a}, and not a disrupted asteroid \citep[{\it cf}.][]{jew10}.
That study further found a mass loss rate roughly an order of magnitude
larger than that calculated for the first known MBC, 133P/Elst-Pizarro 
\citep{hsi04}.  This higher rate of dust emission may
indicate that 238P's activity was triggered much more recently than that of 133P, and therefore is driven by the sublimation
of a much less depleted volatile supply \citep{hsi09a}.

\citet{hag09} found 238P to be dynamically unstable on a time scale of $\sim$20~Myr.
Its proximity to the Themis collisional family, however, suggests that it may have once been
a stable Themis family member, but was destabilized (via a gradual increase of its orbital
eccentricity) by the nearby 1:2 mean-motion
resonance with Jupiter.  As such, despite its current instability, 238P may be physically related, and therefore
compositionally similar, to actual Themis family members and fellow MBCs 133P and 176P/LINEAR.

\section{OBSERVATIONS}

$R$-band observations of 238P were made in photometric conditions between July and December 2010
using the 4.2~m Southern Astrophysical Research (SOAR) telescope at Cerro Pachon and
the 3.54~m New Technology Telescope (NTT) at La Silla, both in Chile, and the 10~m
Keck I and 2.2~m University of Hawaii (UH) telescopes on Mauna Kea in Hawaii.  Details of these
observations are listed in Table~\ref{obslog}.
Observations with SOAR were made using the SOAR Optical Imager \citep[SOI;][]{sch04} which utilizes
a mosaic of two E2V $2048\times4096$~pixel CCDs ($0\farcs154$~pixel$^{-1}$ using 2$\times$2 binning)
behind standard Kron-Cousins broadband filters.
NTT observations were made using the ESO Faint Object Spectrograph and Camera \citep[EFOSC2;][]{buz84},
which employs a $2048\times2048$~pixel Loral/Lesser CCD ($0\farcs24$~pixel$^{-1}$ using 2$\times$2 binning),
behind Bessel broadband filters.
Observations with Keck were made using the Low Resolution Imaging
Spectrometer \citep[LRIS;][]{oke95} in imaging mode.  LRIS employs a
Tektronix $2048\times2048$ CCD with an image scale of
$0\farcs210$~pixel$^{-1}$ and Kron-Cousins filters.
UH 2.2 m observations were made using a Tektronix $2048\times2048$~pixel CCD ($0\farcs219$ pixel$^{-1}$)
behind Kron-Cousins filters.

Standard bias subtraction and flat-field reduction
were performed on all data.  For SOAR, NTT, and UH~2.2~m data, flat
fields were constructed from dithered images of the twilight sky, while
images of the illuminated interior of the dome were used to construct
flat field images for Keck data.  Photometry of \citet{lan92} standard
stars and field stars was obtained by measuring net fluxes within circular
apertures, with background sampled from surrounding circular annuli.
Comet photometry was performed using circular apertures with radii
selected to match the seeing conditions of each night, but to avoid the contaminating
effects of the coma, background sky statistics were measured manually in
regions of blank sky near, but not adjacent, to the object.  Field stars in each
comet image were also measured to assess and correct for
extinction variation during each night.

\section{RESULTS\label{results}}

\subsection{Morphology\label{morphology}}

\begin{figure}
\includegraphics[width=7.0in]{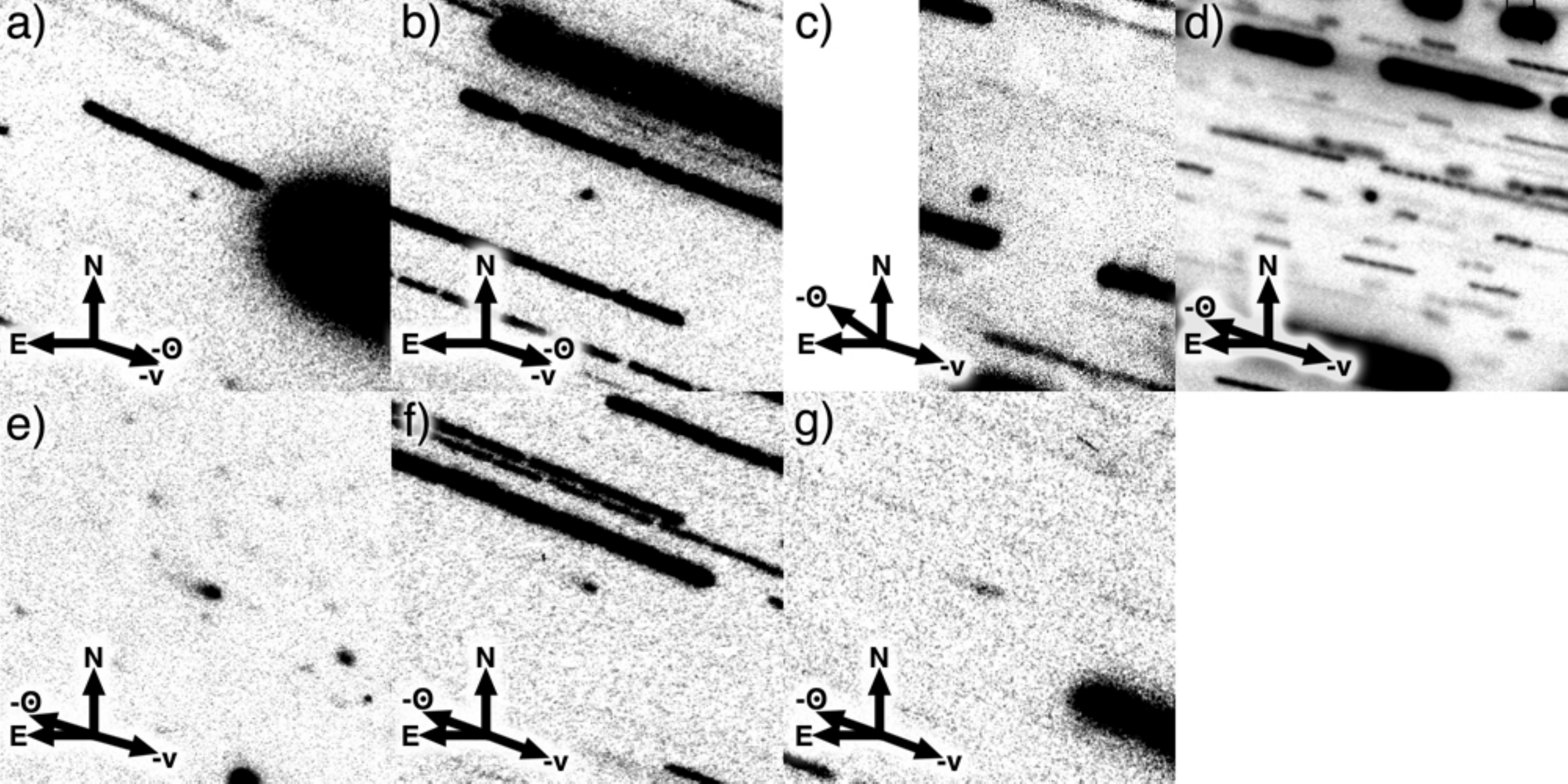}
\caption{\small Composite $R$-band images of 238P (at the center of each panel) constructed from data obtained on
(a) UT 2010 July 7 (8100~s of effective exposure time on the UH 2.2~m),
(b) UT 2010 July 20 (19530~s on the UH 2.2~m),
(c) UT 2010 August 14 (4050~s on SOAR),
(d) UT 2010 September 03 and 04 (23100~s on the NTT),
(e) UT 2010 October 05 (840~s on Keck),
(f) UT 2010 November 25 (4800~s on the UH 2.2~m), and 
(g) UT 2010 December 09 (4500~s on the UH 2.2~m).
All panels are $60\farcs0\times60\farcs0$ in size, with North (N), East (E),
the antisolar direction ($-\odot$), and the negative heliocentric velocity vector ($-v$), as projected
on the sky, marked.
}
\label{images}
\end{figure}

\begin{figure}
\includegraphics[width=6.5in]{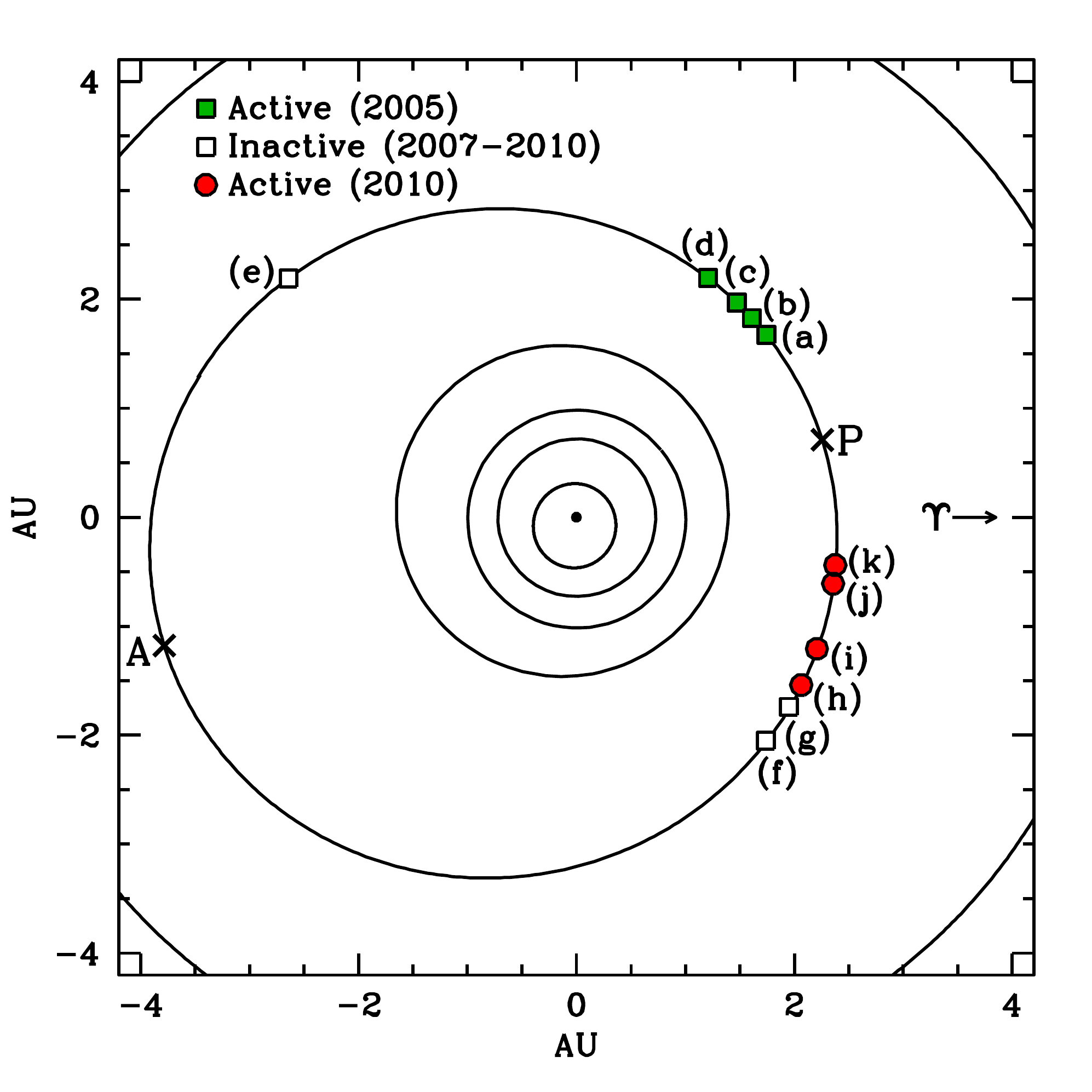}
\caption{\small 
Orbital position plot for 238P observations detailed in Table 1 and in \citet{hsi09a}.  The Sun is shown at the center as a solid
dot, with the orbits of Mercury, Venus, Earth, Mars, 238P and Jupiter (from the center of the plot outwards) shown as black lines.
Green squares mark positions where 238P was observed to be active in 2005 and red circles mark active positions in 2010.
Open squares mark positions where no activity was detected between 2007 and 2010.  
Perihelion (P) and aphelion (A) positions are marked with crosses.  Observations plotted are from
(a) 2005 October 24 \citep{rea05}, (b) 2005 November 10,
(c) 2005 November 19-22, (d) 2005 December 24-25, (e) 2007 January 27, (f) 2010 July 07-20, (g) 2010 August 15,
(h) 2010 September 03-05, (i) 2010 October 05, (j) 2010 November 25, and (k) 2010 December 09, where (b)-(e) are from \citet{hsi09a} and
(f)-(k) are from this work.
}
\label{orbitplot}
\end{figure}

To search for and characterize faint activity for 238P, we obtained multiple exposures of 
the comet each night.  Nightly images were then shifted
(using linear interpolation) to place the comet in the same location and then added
together to create deep composite images (Fig.~\ref{images}).
These deep composite images chronicle a rise in activity for 238P over the course
of these observations.  While in principle, residual dust could remain from
the comet's 2005 outburst, there is a clear transition from a nucleus-dominated morphology
in July, August, and September (Fig.~\ref{images}a-c) to a coma-dominated morphology
in October and November (Fig.~\ref{images}d-e),
the latter of which strongly resembles the morphology observed in 2005
\citep{hsi09a}.  As shown in Figure~\ref{orbitplot}, these observations tightly constrain
the time and orbital position of the onset of 238P's current active episode.
This morphological transition is a strong indication that renewed dust
production took place during this period, a key finding for ascertaining the origin of
that dust production, and an issue we consider further in \S\ref{phsanganalysis}.

\subsection{Photometry\label{photom}}

\subsubsection{Phase Function Analysis\label{phsanganalysis}}

The variation of an object's brightness with solar phase angle (i.e., its phase function)
is dependent on properties of that object's regolith, including 
albedo, particle size distributions, surface roughness, and porosity
\citep[{\it cf}.][]{hel89,mui02}.
Comparisons of phase function parameters found for different objects therefore provide a means
for comparing the objects' surface properties themselves.

\begin{figure}
\includegraphics[width=4.5in]{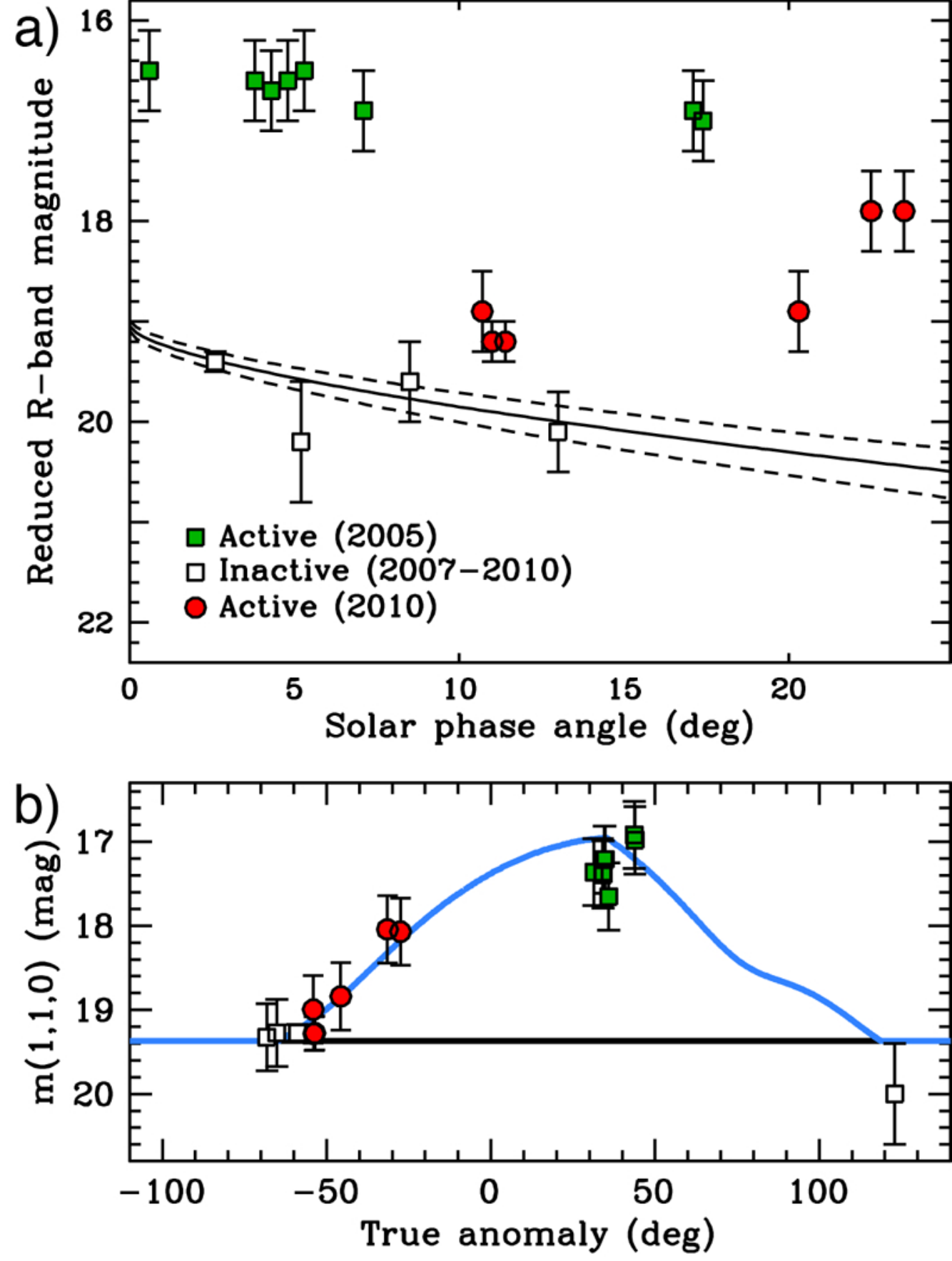}
\caption{\small 
(a) Best-fit phase function (solid line) for reduced $R$-band magnitude data (normalized to $R=\Delta=1$~AU)
from July-August 2010 (open squares) when no activity was detected for 238P (January 2007 photometry is also plotted for reference)
with the range of uncertainty due to phase function parameter uncertainties shown by dotted lines. Data from November-December 2005
\citep[green squares;][]{hsi09a}
and September-December 2010 (red circles) when 238P was active are overplotted.
(b) Plot of results of sublimation modeling where photometry points from (a) are normalized to $\alpha=0\degr$ and to
fixed photometry apertures of $5\farcs0$ at $R=\Delta=1$~AU (assuming a $r^{-1}$ surface brightness profile during active periods)
and replotted, the baseline magnitude of the nucleus is marked by a thick black line (where the phase function is approximated
as linear), and the brightness inside a $5\farcs0$ photometry
aperture predicted by our sublimation model is marked by a thick blue line.
Error bars in both panels indicate the range of uncertainty due to
unknown rotational phases and standard photometric uncertainty,
assuming a peak-to-trough photometric range of $\Delta m\sim0.8$~mag. Given the large uncertainty associated with our January 2007
photometry (at $\nu=123\degr$), it is not explicitly used to fit our model, though we do assume the comet to be inactive at that point.
}
\label{phscurv}
\end{figure}

Using data
obtained in July and August 2010 when 238P was observed to be largely inactive,
we obtain the best-fit phase function plotted in Figure~\ref{phscurv}a, defined by
an absolute magnitude (at $\alpha=0\degr$ and $R=\Delta=1$~AU) of $H=19.05\pm0.07$~mag
and slope parameter $G=-0.03\pm0.10$ in the IAU $H,G$ system (corresponding to a linear phase-darkening coefficient
of $\beta\sim0.045\pm0.007$~mag~deg$^{-1}$).
For this fit, we omit the inactive but extremely faint
January 2007 observation from \citet{hsi09a} which we judged to be unreliable.
For comparison, the slope parameters calculated for MBCs 133P and 176P are $G=0.04\pm0.05$ and $G=0.26\pm0.05$ \citep{hsi09b}, respectively,
and the range of slope parameters
measured for other kilometer-scale Themis asteroids is $-0.23<G<0.60$ \citep{hsi08}.  This result therefore reinforces findings to date
that aside from their cometary activity, MBCs exhibit no observable physical differences from dynamically similar but
inert members of the background asteroid population.
Using a geometric $R$-band albedo of $p_R=0.05$, consistent with that of other MBCs \citep{hsi09b}, we estimate
an effective nucleus radius for 238P of $r_{n}\approx0.4$~km.

Comparison of the remainder of our 2010 data to magnitudes predicted for an inactive 238P nucleus using the phase function
derived here reveals a steadily increasing photometric
excess for data from September through December 2010,
corroborating the conclusion of our morphological analysis (\S\ref{morphology}) that 238P experienced renewed
dust production, likely to be sublimation-driven (see \S\ref{sublmodel}), during this period.
Plotting data (corrected to normalized photometry apertures) from both 2005-2007 and 2010 (Table~\ref{photresults})
as a function of true anomaly (Fig.~\ref{phscurv}b), we find comparable photometric excesses in both data sets, with the excess
observed in 2010 appearing poised to rise to the level observed in 2005.  Further observations will be required
when 238P reaches a comparable point in its orbit in late 2011 to confirm whether its current activity actually 
reaches a level comparable to its previous active episode.

\subsubsection{Sublimation Modeling\label{sublmodel}}

To further analyze the photometric excesses detected in our data (\S\ref{phsanganalysis}), we employ a thermal sublimation
model to estimate the grain flux, which can then be used to derive the estimated photometric enhancement
expected from a sublimating surface \citep{mee86}.  This model is capable of simulating multiple volatile components,
but previous thermal modeling has shown that crystalline water ice is the
only cometary volatile that is likely to be present on MBCs \citep{sch08,pri09}. As such, for 238P,
we only consider the action of H$_2$O ice sublimation.

For the purposes of the model, we assume that the comet's activity profile is identical during each apparition,
permitting us to combine observations from 2005 and 2010 by parameterizing them by true anomaly.
Due to the highly underconstrained nature of this modeling effort, we do not conduct a systematic search of the entire available
parameter space, but instead seek to simply find a plausible sublimation scenario that reproduces our limited observations.
Using standard assumptions for albedo and emissivity used for other comets and primitive asteroids, we find that our data
fit plausibly well to a model in which activity turns on at $\nu=295\degr$ (corresponding to 06 December 2004, or 21 July 2010), rising
exponentially until
reaching a maximum (scaled to match the maximum activity intensity) at $\nu=35\degr$ (24 November 2005, or 06 July 2011),
and then immediately declining exponentially and ceasing at $\nu=120\degr$ (07 January 2007, or 08 August 2012).

Given the numerous
underconstrained free parameters in our model, these activity initiation, peak, and cessation points certainly do not represent a
unique solution and therefore do not necessarily reflect the precise reality of this comet's active behavior. They do, however,
represent a plausible scenario in which 238P's activity is driven by the sublimation of water ice, accomplishing the primary
goal of our modeling efforts.

From our models, we find a likely active area equivalent to $\sim1$\% of 238P's total surface area, which for
a $r_n\approx0.4$~km body, represents an active area of $\sim2\times10^4$~m$^2$, equivalent
to a circular region with a radius of $\sim80$~m.
This radius is about an order of magnitude larger than the
radius of the active area estimated for 133P \citep{hsi04}.  Along with 238P's smaller size and therefore
lower escape velocity for emitted dust particles, the larger physical size of its active area is consistent with its
dramatically more vigorous activity compared to 133P.

Using the models, we also derive a maximum launchable dust grain size of $\sim10$~mm, assuming a freely-sublimating surface.
If sublimation is concentrated in vents, even larger grains could potentially be emitted.
As such, long-lasting residual activity due to slow-dissipating large grains from 238P's previous active apparition in 2005
cannot be ruled out by the results of this modeling.  Whether remnants of previous dust emission episodes are present or not,
however, we already find strong evidence of renewed dust production in analyses discussed in \S\ref{morphology} and \S\ref{phsanganalysis}.

The delay in peak activity for 238P until $\sim35\degr$ past perihelion as implied by the best-fit model
suggests either (1) the ice reservoir driving the observed cometary activity is located at some depth below the surface,
delaying the onset of activity until a solar radiation-driven thermal wave of sufficient intensity is able to propagate
to that depth, or (2) activity is modulated by seasonal effects as hypothesized for 
133P \citep{hsi04} where maximum solar insolation of an isolated active site is primarily a function of an object's obliquity
and solstice position, not heliocentric distance.  If (1) is correct, we would expect all MBCs to exhibit peak activity shortly
after perihelion.  If (2) is correct, we would expect other MBCs to be equally likely to exhibit peak activity before perihelion as well.
Thus far, all the known MBCs primarily exhibit post-perihelion activity, though better characterization of precise activation and
cessation points, as well as the discovery and characterization of a larger sample MBCs, are required to
determine whether (1) or (2) is the primary mechanism for MBC activity modulation. 

\section{DISCUSSION\label{discussion}}

Our finding of renewed dust production 
is strong confirmation that 238P is exhibiting genuine cometary activity
driven by the sublimation of volatile ice.  As concluded for 133P \citep{hsi04}, repeated activity
is extremely unlikely to occur if 238P's observed dust emission is simply an ejecta cloud produced by an impactor striking inert
asteroidal regolith.  This scenario would require two separate collisions to occur on 238P within less than 5 years,
when collisions causing similar behavior are not observed anywhere nearly as frequently on other nearby asteroids.
The extraordinarily coincidental timing required to produce two active episodes at nearly the same portions of 238P's orbit
further decreases the plausibility of impact generation for its activity.
A far more plausible explanation is that 238P's activity is driven by the sublimation of volatile ice,
perhaps preserved in a subsurface reservoir still reachable (after a delay) by solar insolation, or contained
within an isolated area exposed on the surface via collisional excavation and subsequently
subjected to seasonal variations (either due to the object's obliquity, in direct analogy to Earth's seasons, or
possibly also related to the local topography surrounding the active site) in solar radiation.

Observations of 238P's activity in 2010 are thus far insufficient to evaluate
the suggestion made by \citet{hsi09a} that we had 
perhaps witnessed the immediate aftermath of a collisional activation of the MBC in 2005 and that
subsequent outbursts may be much weaker.
The observations of 238P reported here were made at a different portion of its orbit than in 2005 and it is unclear
whether the activity will increase or decline in intensity
as the object progresses to that later portion of its orbit again.  For this reason, further
observations of 238P are highly encouraged when it again becomes observable from Earth in mid-2011.

Even if activity levels during the current active episode remain comparable to those during 2005, it is still
possible that 238P was activated much more recently than less vigorously-active MBCs like 133P and 176P, except that
its activity declines on a more gradual basis than from one orbit to the next.  Long-term monitoring of 238P
(and other vigorously-active MBCs such as P/2008 R1 (Garradd) and P/2010 R2 (La Sagra)) will help clarify this behavior
and likely give valuable insights into the scale of volatile reservoirs in active MBCs, and by extension, in
compositionally similar but otherwise inactive main-belt asteroids, as well as their rate of depletion once excavated.

We find that in comparison to 133P, whose periodic cometary activity appears to begin shortly before
perihelion \citep{hsi10}, 238P shows periodic activity beginning much earlier relative to perihelion.
This discrepancy is consistent with the seasonal hypothesis proposed for the modulation of activity in 133P
(and by extension, in other MBCs) \citep{hsi04}, as it shows that MBC activity is not strictly correlated
to heliocentric distance.  This interpretation is not conclusive proof of this hypothesis, but
does highlight the need for regular monitoring of other MBCs.  Should MBCs consistently show activity only
near perihelion, as most of the currently known MBCs do, this would be a sign that sublimation-driven cometary
activity in these bodies is in fact strongly dependent on heliocentric distance, perhaps subject to moderate delays in the
propagation of solar thermal waves introduced by insulating surface material.  The collisional activation model
for MBCs and correlated seasonal modulation hypothesis suggest, however, that
other MBCs may exhibit activity well before perihelion, since under those
hypotheses, activity modulation depends more on shadowing due to obliquity and topography
than heliocentric distance (though some observational bias towards detections of active MBCs at smaller geocentric,
and therefore heliocentric, distances is expected).

\begin{acknowledgements}
HHH is supported by NASA through Hubble Fellowship grant HF-51274.01 awarded by the Space Telescope Science Institute,
which is operated by the Association of Universities for Research in Astronomy, Inc., for NASA,
under contract NAS 5-26555, and by the UK's Science and Technology Facilities Council (STFC)
through STFC fellowship grant ST/F011016/1.  
Some material presented here was also based on work supported by NASA through the
NASA Astrobiology Institute under Cooperative Agreement 
No. NNA09DA77A issued through the Office of Space Science, and in part through
NASA grant NNX07AF79G.
Additional observing travel support was provided by
the STFC through the Panel for the Allocation of Telescope Time, and the European Southern Observatory.
We are grateful to Sergio Pizarro, Daniel Maturana, Angelica Le\'on, Greg Wirth,
Heather Hershley, John Dvorak, Dave Brennen, Richard Morriarty, and Callie McNew for their assistance with our observations,
and to Bin Yang, Pedro Lacerda, Gal Sarid, and Zahed Wahhaj  for helpful comments on this manuscript.
\end{acknowledgements}

\begin{deluxetable}{llccrrrrrrrc}
\tablewidth{0pt}
\tablecaption{Observation Log\label{obslog}}
\tablecolumns{12}
\tablehead{
\colhead{UT Date}
  & \colhead{Tel.}
  & \colhead{Moon\tablenotemark{a}}
  & \colhead{$\theta_s$\tablenotemark{b}}
  & \colhead{N\tablenotemark{c}}
  & \colhead{$R$\tablenotemark{d}}
  & \colhead{$\Delta$\tablenotemark{e}}
  & \colhead{$\alpha$\tablenotemark{f}}
  & \colhead{$\nu$\tablenotemark{g}}
  & \colhead{$PA_{-\odot}$\tablenotemark{h}}
  & \colhead{$PA_{-v}$\tablenotemark{i}}
  & \colhead{Active?\tablenotemark{j}}
}
\startdata
2010 Jul 07 & UH2.2  & N$-$5  & 0$\farcs$7 & 16 & 2.704 & 1.821 & 13.0 & 291.8 & 253.5 & 251.0 & no \\
2010 Jul 20 & UH2.2  & N$+$8  & 0$\farcs$7 & 23 & 2.674 & 1.709 &  8.5 & 294.8 & 256.0 & 251.4 & no \\
2010 Aug 15 & SOAR   & N$+$5  & 1$\farcs$0 & 21 & 2.616 & 1.608 &  2.6 & 301.1 &  55.0 & 252.9 & no \\
2010 Sep 03 & UH2.2  & N$-$5  & 0$\farcs$7 &  5 & 2.576 & 1.643 & 10.7 & 305.9 &  69.9 & 254.0 & no \\
2010 Sep 04 & NTT    & N$-$4  & 1$\farcs$1 & 54 & 2.574 & 1.647 & 11.0 & 306.1 &  70.1 & 254.1 & yes \\
2010 Sep 05 & NTT    & N$-$3  & 0$\farcs$8 & 35 & 2.572 & 1.651 & 11.4 & 306.4 &  70.3 & 254.1 & yes \\
2010 Oct 05 & Keck   & N$-$2  & 0$\farcs$9 &  8 & 2.514 & 1.869 & 20.3 & 314.3 &  73.0 & 254.4 & yes \\
2010 Nov 25 & UH2.2  & N$-$11 & 0$\farcs$8 &  8 & 2.433 & 2.414 & 23.5 & 328.5 &  71.3 & 251.0 & yes \\
2010 Dec 09 & UH2.2  & N$+$3  & 0$\farcs$8 & 10 & 2.416 & 2.566 & 22.5 & 332.5 &  70.3 & 249.7 & yes
\enddata
\tablenotetext{a} {Phase of moon, in offset from new moon (``N'') in days}
\tablenotetext{b} {Average seeing (FWHM) in arcsec}
\tablenotetext{c} {Number of images}
\tablenotetext{d} {Heliocentric distance (AU)}
\tablenotetext{e} {Geocentric distance (AU)}
\tablenotetext{f} {Solar phase angle (Sun-238P-Earth) in degrees}
\tablenotetext{g} {True anomaly in degrees}
\tablenotetext{h} {Position angle of the antisolar vector, as projected in the plane of the sky, in degrees east of north}
\tablenotetext{i} {Position angle of the negative velocity vector, as projected in the plane of the sky, in degrees east of north}
\tablenotetext{j} {Is visible activity detected?}
\end{deluxetable}

\begin{deluxetable}{lcccrcc}
\tablewidth{0pt}
\tablecaption{Photometric Analysis\label{photresults}}
\tablecolumns{7}
\tablehead{
\colhead{UT Date}
  & \colhead{$m_{R,avg}$\tablenotemark{a}}
  & \colhead{$m_{R,mid}$\tablenotemark{b}}
  & \colhead{$m_R(1,1,\alpha)$\tablenotemark{c}}
  & \colhead{$\delta m_R$\tablenotemark{d}}
  & \colhead{$A_{d}/A_{n}$\tablenotemark{e}}
  & \colhead{$M_{d}$\tablenotemark{f}}
}
\startdata                                                                                                            
2005 Nov 10\tablenotemark{g} & 19.28$\pm$0.05 &     ---      & 16.5$\pm$0.4 &    2.7$\pm$0.4 & 12$\pm$4     & 10$\pm$4 \\ 
2005 Nov 19\tablenotemark{g} & 19.34$\pm$0.05 &     ---      & 16.6$\pm$0.4 &    2.9$\pm$0.4 & 15$\pm$4     & 13$\pm$4 \\ 
2005 Nov 20\tablenotemark{g} & 19.46$\pm$0.05 &     ---      & 16.7$\pm$0.4 &    2.8$\pm$0.4 & 13$\pm$4     & 11$\pm$4 \\ 
2005 Nov 21\tablenotemark{g} & 19.37$\pm$0.05 &     ---      & 16.6$\pm$0.4 &    3.0$\pm$0.4 & 16$\pm$5     & 14$\pm$4 \\ 
2005 Nov 22\tablenotemark{g} & 19.28$\pm$0.05 &     ---      & 16.5$\pm$0.4 &    3.1$\pm$0.4 & 17$\pm$5     & 15$\pm$4 \\ 
2005 Nov 26\tablenotemark{g} & 19.72$\pm$0.05 &     ---      & 16.9$\pm$0.4 &    2.8$\pm$0.4 & 13$\pm$4     & 11$\pm$4 \\ 
2005 Dec 24\tablenotemark{g} & 20.12$\pm$0.03 &     ---      & 16.9$\pm$0.4 &    3.3$\pm$0.4 & 21$\pm$6     & 18$\pm$5 \\ 
2005 Dec 25\tablenotemark{g} & 20.16$\pm$0.03 &     ---      & 17.0$\pm$0.4 &    3.2$\pm$0.4 & 19$\pm$6     & 17$\pm$5 \\ 
2007 Jan 27\tablenotemark{g} & 24.9$\pm$0.4   &     ---      & 20.2$\pm$0.6 & $-$0.6$\pm$0.6 & ---          & --- \\
2010 Jul 07                  & 23.61$\pm$0.10 & 23.6$\pm$0.4 & 20.1$\pm$0.4 & $-$0.1$\pm$0.4 & ---          & --- \\
2010 Jul 20                  & 22.85$\pm$0.06 & 22.9$\pm$0.4 & 19.6$\pm$0.4 &    0.2$\pm$0.4 & ---          & --- \\
2010 Aug 15                  & 22.34$\pm$0.05 & 22.5$\pm$0.1 & 19.4$\pm$0.1 &    0.0$\pm$0.1 & ---          & --- \\
2010 Sep 03                  & 21.97$\pm$0.04 & 22.0$\pm$0.4 & 18.9$\pm$0.4 &    1.0$\pm$0.4 &  2.5$\pm$1.7 &  2.2$\pm$1.5 \\ 
2010 Sep 04                  & 22.01$\pm$0.05 & 22.3$\pm$0.2 & 19.2$\pm$0.2 &    0.7$\pm$0.2 &  1.9$\pm$0.3 &  1.6$\pm$0.3 \\ 
2010 Sep 05                  & 22.02$\pm$0.05 & 22.3$\pm$0.2 & 19.2$\pm$0.2 &    0.7$\pm$0.2 &  1.9$\pm$0.3 &  1.6$\pm$0.3 \\ 
2010 Oct 05                  & 22.25$\pm$0.05 & 22.3$\pm$0.4 & 18.9$\pm$0.4 &    1.4$\pm$0.4 &  3.6$\pm$1.1 &  3.1$\pm$1.0 \\ 
2010 Nov 25                  & 21.75$\pm$0.05 & 21.8$\pm$0.4 & 17.9$\pm$0.4 &    2.5$\pm$0.4 & 10$\pm$3     &  8.7$\pm$2.6 \\ 
2010 Dec 09                  & 21.86$\pm$0.07 & 21.9$\pm$0.4 & 17.9$\pm$0.4 &    2.5$\pm$0.4 & 10$\pm$3     &  8.7$\pm$2.6    
\enddata
\tablenotetext{a} {Mean $R$-band magnitude (averaged in flux space), measured in $4\farcs0$ radius apertures}
\tablenotetext{b} {Estimated midpoint of $R$-band lightcurve, assuming peak-to-trough photometric range of $\Delta m\sim0.8$~mag}
\tablenotetext{c} {Reduced midpoint $R$-band magnitude (normalized to $R=\Delta=1$~AU)}
\tablenotetext{d} {Deviation between measured $m_R(1,1,\alpha)$ and predicted magnitude from best-fit phase function based on
                    data from July-August 2010 when 238P was observed to be inactive}
\tablenotetext{e} {Inferred ratio of scattering surface area of dust to nucleus scattering surface area}
\tablenotetext{f} {Estimated dust mass, in 10$^{4}$ kg, assuming 10 $\mu$m-radius grains and $\rho$=1300~kg~m$^{-3}$, as
                    for 133P \citep{hsi10}, contained within $4\farcs0$ photometry apertures}
\tablenotetext{g} {From \citet{hsi09a}.}
\end{deluxetable}

\end{document}